\documentclass[3p,10pt,fleqn]{elsarticle}

\usepackage{graphicx}
\usepackage[figuresright]{rotating}
\usepackage{bm,amsmath,amssymb}
\usepackage[mathscr]{eucal}
\usepackage{color}
\usepackage{caption}

\biboptions{square,comma,numbers,sort&compress}

\long\def\comment#1{ }
\newcommand{\eqn}[1]{Eq.~\eqref{#1}}
\newcommand{\beq}{\begin{equation}}
\newcommand{\eeq}{\end{equation}}
\newcommand{\nn}{\nonumber\\}
\newcommand{\dif}{{\rm d}}

\newcommand{\rmJ}{{\rm J}}

\newcommand{\order}[1]{\mcal{O}{(#1)}}
\newcommand{\mcal}{\mathcal}

\newcommand{\bx}{\bm{x}}
\newcommand{\by}{\bm{y}}
\newcommand{\bu}{\bm{u}}

\newcommand{\bz}{\bm{z}}

\newcommand{\br}{\bm{r}}

\newcommand{\abar}{\bar{\alpha}_s}

\newcommand{\sdla}{{\rm \scriptscriptstyle DLA}}

\newcommand{\slog}{{\rm \scriptscriptstyle STL}}

\newcommand{\Nc}{N_{\rm c}}
\newcommand{\CF}{C_{\rm F}}
\newcommand{\Nf}{N_{\rm f}}

\newcommand{\minus}{\!-\!}

\begin{document}

\begin{frontmatter}

\title{{\bf Collinearly-improved BK evolution meets the HERA data}}

\author[sac]{E.~Iancu\corref{cor1}}
\ead{edmond.iancu@cea.fr}

\author[sac]{J.D.~Madrigal}
\ead{jose-daniel.madrigal-martinez@cea.fr}

\author[col]{A.H.~Mueller}
\ead{amh@phys.columbia.edu}

\author[sac]{G.~Soyez}
\ead{gregory.soyez@cea.fr}

\author[ect]{D.N.~Triantafyllopoulos}
\ead{trianta@ectstar.eu}

\address[sac]{Institut de Physique Th\'{e}orique, CEA Saclay, CNRS UMR 3681, F-91191 Gif-sur-Yvette, France}
\address[col]{Department of Physics, Columbia University, New York, NY 10027, USA}
\address[ect]{European Centre for Theoretical Studies in Nuclear Physics and Related Areas (ECT*)\\and Fondazione Bruno Kessler, Strada delle Tabarelle 286, I-38123 Villazzano (TN), Italy}

\cortext[cor1]{Corresponding author}

\begin{abstract}
In a previous publication, we have established a collinearly-improved version of the Balitsky-Kovchegov (BK) equation,
which resums to all orders the radiative corrections enhanced by large double transverse logarithms. Here, we study
the relevance of this equation as a tool for phenomenology, by confronting it to the HERA data. To that aim, we
first improve the perturbative accuracy of our resummation, by including two classes of single-logarithmic corrections:
those generated by the first non-singular terms in the DGLAP splitting functions and those expressing the
one-loop running of the QCD coupling. The equation thus obtained includes 
all the next-to-leading order corrections to the BK equation which are enhanced by (single or double) collinear
logarithms. We then use numerical solutions to this equation to fit the HERA data for the electron-proton reduced
cross-section at small Bjorken $x$. We obtain good quality fits for physically acceptable initial conditions.
Our best fit, which shows a good stability up to virtualities as large as $Q^2=400\,\text{GeV}^2$ for the exchanged
photon, uses as an initial condition the running-coupling version of the McLerran-Venugopalan model,
with the QCD coupling running according to the smallest dipole prescription.

\end{abstract}


\end{frontmatter}

\section{Introduction}
\label{sect:intro}
The wealth of data on electron-proton deep inelastic scattering collected by the experiments at HERA over
15 years of operation has allowed for stringent tests of our understanding of high-energy scattering from first
principles. This refers in particular to the `small-$x$' regime where perturbative QCD predicts a rapid growth of the
gluon density with increasing energy (or decreasing Bjorken $x$), 
leading to non-linear phenomena like multiple scattering
and gluon saturation \cite{Gelis:2010nm,Kovchegov:2012mbw}. The simplicity
of the dipole factorization for deep inelastic scattering at high energy 
\cite{Mueller:1989st,Nikolaev:1990ja} has favored the emergence of relatively 
simple `dipole models', in which the high-density effects are efficiently implemented as unitarity 
corrections to the cross-section for the scattering between a quark-antiquark dipole and the proton. 
Such models allowed for rather successful fits to the small-$x$ HERA data
 at a time where the theory 
of the non-linear evolution in QCD was insufficiently developed and the pertinence of gluon saturation for 
the phenomenology was far from being widely accepted. The first such model
--- the ``GBW saturation model'' \cite{GolecBiernat:1998js,GolecBiernat:1999qd}
--- provided a rather good description of the early HERA data for the inclusive and diffractive structure
functions at  $x\le 10^{-2}$ with only 3  free parameters. This success
 inspired new ways to look at the HERA data, which in particular led to the 
identification of geometric scaling  \cite{Stasto:2000er}. The subsequent understanding
\cite{Iancu:2002tr,Mueller:2002zm,Munier:2003vc} of this scaling from the non-linear evolution equations 
in QCD --- the Balitsky-JIMWLK hierarchy
  \cite{Balitsky:1995ub,JalilianMarian:1997jx,JalilianMarian:1997gr,Kovner:2000pt,Iancu:2000hn,Iancu:2001ad,Ferreiro:2001qy} and its mean field approximation known as the Balitsky-Kovchegov (BK) equation
\cite{Kovchegov:1999yj} --- 
has greatly increased our confidence in the validity of the pQCD approach to gluon saturation as a valuable
tool for phenomenology.

Over the next years, new `dipole models', of increasing
sophistication, have emerged.  On one hand, they were better rooted in
perturbative QCD, thus reflecting the overall progress of the theory
\cite{Bartels:2002cj,Iancu:2003ge,Albacete:2009fh,Albacete:2010sy,Kuokkanen:2011je}.
On the other hand, they were better constrained by the advent of new
data at HERA, of higher precision. Finally, they extended the scope of
the `saturation models' to other observables, like diffractive
structure functions and particle production in heavy ion collisions.
Such extensions required more elaborated versions of the dipole model,
including impact parameter dependence 
\cite{Kowalski:2003hm,Kowalski:2006hc,Rezaeian:2012ji,Rezaeian:2013tka}
and heavy quarks \cite{Kowalski:2006hc,Soyez:2007kg}.

For several years, the theory of high-energy scattering with high gluon density was known only 
to leading logarithmic accuracy in pQCD, which is insufficient for direct applications 
to phenomenology. For instance, the essential running coupling corrections enter the high energy evolution
only at next-to-leading order (see below for details).  
To cope with that, the first generations of dipole models involved phenomenological
parametrizations for the dipole amplitude, which were rather ad-hoc, albeit sometimes inspired by solutions 
to the BK-JIMWLK equations. For instance, the `IIM' fit in \cite{Iancu:2003ge} attempted to capture some 
general features of the non-linear evolution, like geometric scaling with an anomalous 
dimension and the BFKL diffusion, that were expected to hold beyond leading order 
\cite{Mueller:2002zm,Triantafyllopoulos:2002nz}. However, the situation has changed in the recent years,
when the next-to-leading corrections to the BK and JIMWLK equations have progressively become available 
\cite{Kovchegov:2006wf,Kovchegov:2006vj,Balitsky:2006wa,Balitsky:2008zza,Balitsky:2013fea,Kovner:2013ona}.
This opened the possibility for new fits in which the evolution of the dipole amplitude with increasing
energy is completely fixed by the theory
and only the initial condition at low energy still requires some
modeling involving free parameters.
In that respect, the situation of modern `dipole fits' becomes
comparable in spirit to that of the more traditional fits based on the
DGLAP equation.

So far, this strategy has been applied 
\cite{Albacete:2009fh,Albacete:2010sy,Kuokkanen:2011je,Lappi:2013zma} only at the level of the 
``running coupling BK equation'' (rcBK) ---  an improved version of the LO BK equation which resums 
all-order corrections associated with the running of coupling, with some scheme dependence though 
\cite{Kovchegov:2006wf,Kovchegov:2006vj,Balitsky:2006wa}.  These corrections are numerically large,
since enhanced by a large transverse (or `collinear') logarithm, 
and their resummation within the BK equation has important consequences on the evolution ---
it significantly slows down the growth of the dipole amplitude with increasing energy
\cite{Mueller:2002zm,Triantafyllopoulos:2002nz,Albacete:2007yr}. This last feature was indeed essential
for the success of the HERA fits based on rcBK \cite{Albacete:2009fh,Albacete:2010sy,Kuokkanen:2011je,Lappi:2013zma}. 
The state of the art in that sense is the ``AAMQS''  fit in \cite{Albacete:2010sy}, which provides
a good description of the most recent HERA data \cite{Aaron:2009aa} (the combined analysis by
H1 and ZEUS for the $ep$ reduced cross-section, which is characterized by very small error bars),
with a number of free parameters which varies from 4 to 7 (depending upon whether heavy quarks
are included in the fit, or not).

However, the running of the QCD coupling is not the only source of large 
(but formally higher-order) perturbative corrections to the LO BK, or
JIMWLK, equations. Besides the running coupling corrections, the
full next-to-leading order (NLO) corrections to the BK equation, as computed in \cite{Balitsky:2008zza},
feature other contributions which are enhanced by potentially large, single or double, transverse
logarithms. Such terms were indeed expected, given our
experience with the NLO version \cite{Fadin:1995xg,Fadin:1997zv,Camici:1996st,Camici:1997ij,Fadin:1998py,Ciafaloni:1998gs}
of the BFKL equation \cite{Lipatov:1976zz,Kuraev:1977fs,Balitsky:1978ic}  (the linearized version of the BK equation
valid when the scattering is weak). The NLO BFKL corrections are numerically large and thus render the small--$x$
evolution, at LO and NLO, void of any predictive power. There is no reason to expect this problem to be cured, or even alleviated, 
by the inclusion of the non-linear terms describing unitarity corrections  \cite{Triantafyllopoulos:2002nz,Avsar:2011ds}: 
the collinear logarithms are generated by integrating over regions in phase-space where
the dipole size is small and the scattering is weak.
This has been indeed confirmed by the first numerical study of the
NLO BK equation  \cite{Lappi:2015fma}, which showed that the evolution is unstable 
(the scattering amplitude decreases with increasing energy and can even turn negative) and that the main
source for such an instability is the large double-logarithmic correction.

This difficulty reflects the existence of large radiative corrections
of higher orders in $\alpha_s$, which formally lie outside the scope
of the high-energy evolution (since generated by the transverse
phase-space), but in practice spoil the convergence of the
perturbation theory and hence must be kept under control via
appropriate resummations.  In a previous publication
\cite{Iancu:2015vea}, we have devised a resummation scheme which deals
with the largest such corrections --- those where each power of
$\alpha_s$ is accompanied by a double transverse logarithm. Our
strategy relies on explicit calculations of Feynman graphs and results
in an effective evolution equation --- a collinearly improved version
of LO BK equation ---, in which both the kernel and the initial
condition receive double-logarithmic corrections to all orders.  This
scheme differs from the `collinear resummations' previously proposed
in the context of NLO BFKL
\cite{Kwiecinski:1997ee,Salam:1998tj,Ciafaloni:1999yw,Ciafaloni:2003rd,Vera:2005jt}
in that it is explicitly formulated in the transverse coordinate
space, rather than in Mellin space, and hence it is consistent with
the non-linear structure of the BK equation. Besides, our equation is
local in `rapidity' (the logarithm of the energy, which plays the role
of the evolution variable), a property which in this context is rather
remarkable since 
the physics behind the double collinear logarithms is the
time-ordering of subsequent, soft, gluon emissions, which is genuinely
non-local.\footnote{In fact, a non-local equation to resum the double
  logarithms has been proposed in \cite{Beuf:2014uia}.} The first
numerical studies of this collinearly-improved BK equation demonstrate
the essential role played by the resummation in both stabilizing and
slowing down the evolution \cite{Iancu:2015vea,Manty-HP2015}.

In this paper, we shall provide the first phenomenological test of our resummation scheme,
by using it in fits to the inclusive HERA data. To that aim, it will be important to first
extend this scheme to also include the {\em single} transverse logarithms
which appear in the NLO correction to the BK equation --- that is, the NLO terms 
expressing the first correction to the DGLAP
splitting kernel beyond the small--$x$ approximation
and those associated with the one-loop running 
of the coupling. Indeed, such single-log effects must be kept under control to
ensure a good convergence of the perturbative expansion. Besides, the inclusion of running coupling 
effects is essential for the description of the data, as well known.

The resummation of the DGLAP logarithms to the order of interest turns out to be rather
straightforward: it amounts to adding an anomalous dimension (a piece of the leading-order
DGLAP anomalous dimension)  to both the resummed kernel and  
the resummed initial condition. For the running coupling corrections, the situation turns out to be more
subtle since, strictly speaking, they cannot be encoded into an equation which is local in 
rapidity. This being said, and following the standard strategy in the literature, we shall propose various
schemes for introducing a running coupling directly in the local evolution equation and test
these schemes via fits to the HERA data.

After these additional resummations, we are led to a new, more refined, version for the
`collinearly improved BK equation', namely \eqn{colbk} below, which will be our main tool
for phenomenology. By construction, this equation resums the double-logarithmic corrections
{\em completely} --- meaning to all orders in $\abar\equiv \alpha_s N_c/\pi$
($\alpha_s$ is the QCD coupling and $N_c$ is the number of colors)
and with the right symmetry factors ---,
whereas the single-logarithmic terms are resummed only {\em partially} (but in such a way
to include the respective terms to NLO).
It is rather straightforward to extend our resummed equation to full NLO accuracy,
by adding the remaining corrections of $\order{\abar^2}$, as computed in \cite{Balitsky:2008zza}.
But the ensuing equation would be very cumbersome to use in practice, due to the
intricate, non-local and non-linear, structure of the pure $\abar^2$ corrections. In this first
analysis, we shall adopt the viewpoint that the most important higher-order contributions
(say, in view of phenomenology) are those enhanced by collinear logs, as explicitly resummed
in \eqn{colbk}, and that the pure $\abar^2$ effects are truly small and can be effectively taken 
care of via the fitting procedure. A similar viewpoint has been advocated in previous fits based on 
rcBK, but given the importance of the collinear logarithms,
this assumption was not so well motivated and led indeed
to some tensions in the respective fits, as we shall later explain.

Using numerical solutions to this collinearly improved BK equation together with suitable forms
for the initial condition, we have performed fits to the HERA data
for the $ep$ reduced cross-section  \cite{Aaron:2009aa} at $x\le 10^{-2}$ and $Q^2\le Q_{\rm max}^2$,
where the upper limit $Q_{\rm max}^2$ on the virtuality $Q^2$ of the exchanged photon is varied between
50~GeV$^2$ (a common choice in small--$x$ fits) and 400\,GeV$^2$.
These fits show several remarkable characteristics.

\texttt{(i)} The fits are indeed successful: for $Q_{\rm max}^2=50\,$GeV$^2$ and two types of
 initial conditions --- GBW--like \cite{GolecBiernat:1998js} and the 
 running-coupling version of the McLerran-Venugopalan (rcMV) model \cite{McLerran:1993ka} ---,
 we obtain a $\chi^2$ per number of data points around $1.2$ with only 4 free parameters.

\texttt{(ii)} The fits are also very discriminatory: they clearly favor some initial conditions
over some others, and some prescriptions for the running of the coupling over the others.
For instance, the standard MV initial condition, which truly corresponds to a fixed coupling, appears
to be disfavored, whereas a more physical version of it, including a running coupling, works quite well.
The latter works also better than the GBW initial condition, in the sense that it provides
a fit which remains stable up to $Q^2=400\,$GeV$^2$.

\texttt{(iii)} Our fits alleviate some tensions (in terms of physical interpretation) which were visible
in previous first based on rcBK \cite{Albacete:2009fh,Albacete:2010sy,Kuokkanen:2011je,Lappi:2013zma}
and could be attributed to the choice to replace all the NLO 
corrections with the running of the coupling alone (see also the related discussion in \cite{Kuokkanen:2011je}).
Notably,  our fits prefer prescriptions where the QCD coupling $\alpha_s(\mu^2)$ is running according to the
smallest dipole size, they do not require any artificial `anomalous dimension' in the initial condition,
and treat the heavy quarks on the same footing as the light ones, in agreement with general expectations
from the dipole factorization.


\section{The NLO BK equation and large transverse logarithms}
\label{sect:nlobk}

To motivate the resummations that we shall later perform, let us first
explicitly exhibit the large transverse logarithms
which appear when computing the NLO corrections to the BK equation 
\cite{Balitsky:2006wa,Kovchegov:2006vj,Balitsky:2008zza}. We recall that the BK equation
describes the rapidity evolution of the $S$-matrix $S_{\bx\by} = 1 -T_{\bx\by}$
for the scattering of a color dipole with transverse coordinates ($\bx,\by$) off a hadronic target. 
The dipole scattering amplitude $T_{\bx\by}$ is small in the regime where
the target is dilute, but it approaches the unitarity (or `black disk') limit $T_{\bx\by}=1$ when the target
is dense. The separation between these two regimes
is controlled by the saturation momentum $Q_s(Y)$, which increases with the rapidity difference
$Y$ between the projectile and the target.
  
Neglecting the terms suppressed in the limit of a large number of colors $\Nc\gg 1$, one finds
a closed equation for $S_{\bx\by}$, whose NLO version reads as follows \cite{Balitsky:2008zza}
 \begin{align}
 \label{nlobk}
 \hspace*{-0.7cm}
 \frac{\dif S_{\bx\by}}{\dif Y} = \,&
 \frac{\abar}{2 \pi}
 \int \dif^2 \bz \,
 \frac{(\bx\minus\by)^2}{(\bx \minus\bz)^2 (\by \minus \bz)^2}\,
 \bigg\{ 1 + \abar
 \bigg[\bar{b}\, \ln (\bx \minus \by)^2 \mu^2 
 -\bar{b}\,\frac{(\bx \minus\bz)^2 - (\by \minus\bz)^2}{(\bx \minus \by)^2}
 \ln \frac{(\bx \minus\bz)^2}{(\by \minus\bz)^2}
 \nn
 & \hspace*{4.3cm}
 +\frac{67}{36} - \frac{\pi^2}{12} - \frac{5 \Nf}{18 \Nc}- 
 \frac{1}{2}\ln \frac{(\bx \minus\bz)^2}{(\bx \minus\by)^2} \ln \frac{(\by \minus\bz)^2}{(\bx \minus\by)^2}\bigg] 
 \bigg\}
 \left(S_{\bx\bz} S_{\bz\by} - S_{\bx\by} \right)
 \nn
  +\, & \frac{\abar^2}{8\pi^2}
 \int \frac{\dif^2 \bu \,\dif^2 \bz}{(\bu \minus \bz)^4}
 \bigg\{-2
 + \frac{(\bx \minus\bu)^2 (\by \minus\bz)^2 + 
 (\bx \minus \bz)^2 (\by \minus \bu)^2
 - 4 (\bx \minus \by)^2 (\bu \minus \bz)^2}{(\bx \minus \bu)^2 (\by  \minus \bz)^2 - (\bx \minus \bz)^2 (\by \minus \bu)^2}
 \ln \frac{(\bx \minus \bu)^2 (\by  \minus \bz)^2}{(\bx \minus \bz)^2 (\by \minus \bu)^2}
 \nn
 & \hspace*{2.6cm} +
 \frac{(\bx \minus \by)^2 (\bu \minus \bz)^2}{(\bx \minus \bu)^2 (\by  \minus \bz)^2}
 \left[1 + \frac{(\bx \minus \by)^2 (\bu \minus \bz)^2}{(\bx \minus \bu)^2 (\by  \minus \bz)^2 - (\bx \minus \bz)^2 (\by \minus \bu)^2} \right]
 \ln \frac{(\bx \minus \bu)^2 (\by  \minus \bz)^2}{(\bx \minus \bz)^2 (\by \minus \bu)^2}\bigg\}
 \nn
 & \hspace*{2.6cm} \left(S_{\bx\bu} S_{\bu\bz} S_{\bz\by} - S_{\bx \bu} S_{\bu \by}\right)
 \nn
 +\, & \frac{\abar^2}{8\pi^2}\,
 \frac{\Nf}{\Nc}
 \int \frac{\dif^2 \bu \,\dif^2 \bz}{(\bu \minus \bz)^4}
 \bigg[2
 - \frac{(\bx \minus\bu)^2 (\by \minus\bz)^2 + 
 (\bx \minus \bz)^2 (\by \minus \bu)^2
 - (\bx \minus \by)^2 (\bu \minus \bz)^2}{(\bx \minus \bu)^2 (\by  \minus \bz)^2 - (\bx \minus \bz)^2 (\by \minus \bu)^2}
 \ln \frac{(\bx \minus \bu)^2 (\by  \minus \bz)^2}{(\bx \minus \bz)^2 (\by \minus \bu)^2} \bigg]
 \nn
 & \hspace*{2.6cm} \left( S_{\bx\bz} S_{\bu\by}- S_{\bx\bu} S_{\bu\by} \right),
 \end{align}
where $\Nf$ is the number of flavors, $\bar{b} = (11\Nc - 2\Nf)/12 \Nc$, and $\abar = \alpha_s \Nc /\pi$,
with the QCD coupling $\alpha_s$ evaluated at the renormalization scale $\mu$.

There are two main changes in the structure of the evolution equation as we go from LO to NLO. First, the term with a single integration (SI) over the transverse coordinate $\bz$ only receives a correction of order $\mathcal{O}(\abar^2)$
to the kernel, which in particular contains the running coupling corrections proportional to $\bar{b}$. 
Second, there are new terms, of order $\mathcal{O}(\abar^2)$, which involve a double integration (DI) over the 
transverse coordinates $\bu$ and $\bz$ and which refer to partonic fluctuations involving two additional
partons (besides the original quark and antiquark) at the time of scattering.
The first such a term, which is independent of $\Nf$,  represents fluctuations where both daughter
partons are gluons. The $S$-matrix structure therein, that is, $S_{\bx\bu} S_{\bu\bz} S_{\bz\by} - S_{\bx \bu} S_{\bu \by}$,
corresponds to the following sequence of emissions: the original dipole $(\bx,\by)$ emits a gluon at $\bu$,
thus effectively splitting into two dipoles  $(\bx,\bu)$ and  $(\bu,\by)$; then, the dipole $(\bu,\by)$ emits a gluon
at $\bz$, thus giving rise to the dipoles  $(\bu,\bz)$ and  $(\bz,\by)$. The `real' term 
$S_{\bx\bu} S_{\bu\bz} S_{\bz\by}$ describes the situation where both daughter gluons interact with the
target. The `virtual' term $ - S_{\bx \bu} S_{\bu \by}$ describes the case where the gluon at $\bz$
has been emitted and reabsorbed either before, or after, the scattering.
This negative `virtual' term subtracts the equal-point contribution ($\bz=\bu$) from
 the `real' piece, ensuring that the potential `ultraviolet' singularity associated with 
 the factor $1/(\bu - \bz)^4$ in the kernel is truly harmless. A similar discussion applies to the second DI term,
 proportional to $\Nf$, except for the fact that the additional partons at the time of scattering are a quark and
 an antiquark.

In principle, one should be able to undertake the task of solving the
NLO BK equation. The hope would be that the solution would only add a
relatively small correction to the LO result. However, this is not the
case since there are terms in the kernels of the NLO equation which
can become large in certain kinematic regimes and thus invalidate the
strict $\abar$-expansion. One obvious
class of such terms contains the corrections proportional to $\bar b$  in the SI term in \eqn{nlobk}, 
which by themselves bring no serious difficulties: as well known,
these corrections can be absorbed into a redefinition of the scale for the running of the coupling,
which thus becomes a {\em dynamical} scale (see Sect.~\ref{sect:RC} below for details). Here, we would like
to focus on the corrections enhanced by `collinear  logarithms', that is, logarithms associated with the large
separation in transverse sizes (or momenta) between successive emissions. These corrections become
large only in the weak-scattering regime where all the dipoles are small compared to the
saturation scale $1/Q_s(Y)$ and the equation can be linearized w.r.t. to the (small) scattering amplitude $T$. 
This in particular means that one can ignore
the last term, proportional to $\Nf/\Nc$, in \eqn{nlobk} since this term vanishes after linearization, as one can easily check
(by also using the symmetry of the kernel under  the interchange $\bu \leftrightarrow \bz$).

To be more precise, let us consider the strongly ordered regime
\beq
\label{stor}
 1/Q_{s} \gg |\bz-\bx|\simeq |\bz-\by| \simeq |\bz-\bu| \gg |\bu-\bx| \simeq |\bu-\by| \gg |\bx-\by|,
\eeq   
that is, the parent dipole is the smallest one, a gluon is emitted far away at $\bu$, a second one even further at $\bz$,
but with all possible dipole sizes remaining smaller than the inverse
saturation momentum. 
Whenever appropriate, we will denote by $r$, $\bar{u}$ and $\bar{z}$
the size of the parent dipole, the size of the dipoles involving $\bu$
and the size of the dipoles involving $\bz$, respectively, with
$r^2\ll \bar{u}^2\ll \bar{z}^2$.
By inspection of
the SI piece in the NLO BK equation, it is quite obvious that the dominant NLO term is the one involving a 
double transverse logarithm (DTL), that is, the last term within the square brackets. Still within this regime
\eqref{stor}, we can approximate the scattering matrices in the SI term as follows: $S_{\bx\bz} S_{\bz\by} - S_{\bx\by} 
\simeq -T_{\bx\bz} -T_{\bz\by} + T_{\bx\by} \simeq -2 T(\bar{z})$, where the second approximate equality follows
since the dipole amplitude for a small dipole is roughly proportional to the dipole size squared. Notice that the net
result in the approximation of interest fully comes from the `real' term, which involves the large daughter dipoles.

What is not immediately obvious
is the presence of a single transverse logarithm (STL) coming from the DI term. Let us isolate here the relevant part of the kernel,
\beq
 \label{mstl}
 \mathcal{M}_{\slog} \equiv
 \frac{1}{8(\bu \minus \bz)^4}
 \bigg[-2
 + \frac{(\bx \minus\bu)^2 (\by \minus\bz)^2 + 
 (\bx \minus \bz)^2 (\by \minus \bu)^2
 - 4 (\bx \minus \by)^2 (\bu \minus \bz)^2}{(\bx \minus \bu)^2 (\by  \minus \bz)^2 - (\bx \minus \bz)^2 (\by \minus \bu)^2}
 \ln \frac{(\bx \minus \bu)^2 (\by  \minus \bz)^2}{(\bx \minus \bz)^2 (\by \minus \bu)^2} \bigg].
 \eeq 
To implement the limit in \eqn{stor} we can successively write the expression in \eqn{mstl} as
\beq
\label{mstllim}
\mathcal{M}_{\slog} \simeq
 \frac{1}{8\bar{z}^4}
 \bigg[-2 + \frac{2 \bar{u}^2 - 2 \bar{u} r \cos\phi -3 r^2}{r^2 - 2 \bar{u} r \cos\phi}
 \ln \left(1+ \frac{r^2 - 2 \bar{u} r \cos\phi}{\bar{u}^2} \right)
  \bigg]
  \simeq 
 -\frac{6 -\cos^2\phi}{12}\,
 \frac{r^2}{\bar{u}^2\bar{z}^4}
 \to
 -\frac{11}{24}\,
 \frac{r^2}{\bar{u}^2\bar{z}^4},
\eeq  
with $\phi$ the angle between $\br$ and any of the two dipoles
involving $\bu$.
To obtain \eqref{mstllim}, we have first set all dipole sizes which
include $\bz$ equal to each other, since any subleading term would be
suppressed by inverse powers of $\bar{z}$. Then the only $z$
dependence left is the one explicit in the prefactor. We have
subsequently taken the limit $r \ll \bar{u}$ (by expanding the
logarithm to cubic order) and we have finally averaged over the angle
$\phi$ between the parent dipole and those involving $\bu$.  Notice
that the would-be leading term, of order $1/\bar{z}^4$, has cancelled
out in these manipulations.  The first non-vanishing term, as visible
in the r.h.s. of \eqn{mstllim}, is suppressed by $r^2/\bar{u}^2$, thus
creating the conditions for a logarithmic integration over
$\bar{u}$. To explicitly see this, recall that we consider the
weak-scattering regime, where the product of $S$-matrices multiplying
$\mathcal{M}_{\slog}$ can be linearized. This allows us write
$S_{\bx\bu} S_{\bu\bz} S_{\bz\by} - S_{\bx\bu}S_{\bu\by} \simeq
-T_{\bu\bz} - T_{\bz\by} + T_{\bu\by} \simeq -2T(\bar{z})$. (Once
again, the dominant contribution has been generated by the `real'
term.)  We see that the net scattering amplitude in this approximation
is independent of the intermediate dipole size $\bar{u}$.
Accordingly, when integrating over $\bar{u}$, within the range limited
by $r$ and $\bar{z}$, we find a STL, as anticipated.  After also
including the LO term and the NLO one enhanced by the DTL, one finds
that the NLO BK equation in the strongly ordered region \eqref{stor}
reduces to
 \begin{equation}
 \label{dtdylogs}
 	\frac{\dif T(r)}{\dif Y}
 	= \abar \int_{r^2}^{1/Q_s^2}
 	\dif\bar{z}^2\, \frac{r^2}{\bar{z}^4}
 	\left(1 
 	-\frac{1}{2}\,\abar
 	\ln^2 \frac{\bar{z}^2}{r^2} - \frac{11}{12}\,\abar
 	\ln \frac{\bar{z}^2}{r^2} \right) T(\bar{z}).
 \end{equation}
 It is now clear that, if the daughter dipoles are allowed to
 become sufficiently large, the NLO contributions enhanced by large
 transverse logarithms become comparable to, or larger than, the LO
 one. In that case, the present perturbative expansion cannot be
 trusted anymore.
 To be more explicit, consider a single step $\Delta Y$ in the
 evolution with the following, simple but physically meaningful,
 initial condition
 \beq
 T(r) =
 \begin{cases}
 	r^2 Q_s^2 \quad &\mbox{for} \quad 
 	r^2 Q_s^2 \ll 1
 	\\
 	1 \quad &\mbox{for} \quad r^2 Q_s^2 \gg 1.
 \end{cases}
 \eeq    
The $\bar{z}$-integration in \eqn{dtdylogs} becomes logarithmic and gives 
 \begin{equation}
 \label{dtonestep}
 	\Delta T(r) = \abar \Delta Y r^2 Q_s^2\,\ln \frac{1}{r^2 Q_s^2}
 	\left( 1 - \frac{1}{6}\,\abar \ln^2\frac{1}{r^2 Q_s^2}	- \frac{11}{24}\,\abar \ln \frac{1}{r^2 Q_s^2}
 	  \right).
 \end{equation}
This shows that, for sufficiently small $r Q_s$, such that $\abar \ln^2(1/r^2 Q_s^2) \gtrsim 1$, the NLO correction becomes larger than the LO term and the perturbation series 
is unreliable. In particular, the NLO correction is negative rendering the solution unstable,
 as indeed observed in numerical solutions \cite{Avsar:2011ds,Lappi:2015fma,Iancu:2015vea}.

\section{Resumming the large collinear logarithms}
\label{sect:resum}

The large NLO corrections that we have singled out in the previous section are the lowest-order examples
of collinearly-enhanced radiative corrections, which occur to all orders and spoil the convergence of
the perturbation theory. When the separation between the transverse scales of the projectile and
the target is large enough, as is actually the case in the DIS kinematics at HERA, 
the higher-order terms of this type become more important than
the pure $\abar^2$ NLO terms (i.e. the contributions of
$\order{\abar^2}$ which are not amplified by any
transverse logarithms). From now on, we shall focus on this situation, 
discarding the pure $\abar^2$ NLO corrections, but focusing on the resummation of the large transverse 
logarithms to all orders. In this section, we consider both the single and double collinear logarithms, 
thus following and expanding our recent results in  \cite{Iancu:2015vea}. In the next section,
we shall explain how the running coupling corrections can be included in this scheme.

In Ref.~\cite{Iancu:2015vea} we have devised a strategy for resumming
double-logarithmic corrections to the BK equation to all orders. Our
main observation was that these corrections are generated by the
diagrams common to the BFKL and DGLAP evolutions --- i.e. the Feynman
graphs of light-cone perturbation theory in which the successive gluon
emissions are strongly ordered in both longitudinal momenta and
transverse momenta (or `dipole sizes') --- after enforcing the
additional constraint that the emissions must also be ordered in {\em
  lifetimes} (or, equivalently, in light-cone energies \cite{Beuf:2014uia}). Concerning
the single collinear logarithms, it is intuitively clear that they
must represent DGLAP-like corrections to BFKL, `small-$x$', emissions.
For instance, the effect of order $\abar^2\Delta Y\rho^2$ visible in
\eqn{dtonestep} is the result of a sequence of two emissions: one
small-$x$ emission (in the double logarithmic regime) yielding a
contribution $\propto \abar\Delta Y\rho$, and a DGLAP-like emission,
characterized by strong ordering in dipole sizes (see eq. \eqn{stor})
and which gives an effect of order $\abar\rho$. Here, $\rho\equiv
\ln({1}/{Q_s^2r^2})$ is the large logarithm generated by the
transverse phase-space.  This scenario is corroborated by the
following observation: the numerical coefficient $A_1\equiv 11/12$ in
front of the STL in \eqn{dtdylogs} can be recognized as the
second-order term in the small $\omega$ expansion of the relevant
linear combination of DGLAP anomalous dimensions:
\begin{equation}
    \label{pomega}
    P_{\rm T}(\omega) =
 	\int_0^1 \dif z\, z^\omega 
 	\left[ P_{\rm gg}(z) + \frac{\CF}{\Nc}\, P_{\rm qg}(z) \right]
 	= \frac{1}{\omega} - A_1
 	+\mathcal{O}\left(\omega,\frac{\Nf}{\Nc^3}\right)
 	\quad \mbox{with} \quad A_1 = \frac{11}{12}.
 \end{equation} 
Recalling that one needs one factor of $1/\omega$ in order to generate a small-$x$ logarithm
$\Delta Y=\ln(1/x)$, one sees that the NLO effect $\sim\abar^2\Delta Y\rho^2$ is 
indeed produced by combining the singular ($1/\omega$) piece of one emission 
with the first non-singular piece ($A_1$) of another one. This discussion also instructs us
about the strategy to follow in order to resum such STLs to all orders: it suffices to include
this piece $A_1$ as an anomalous dimension, i.e. as an extra power-law suppression, 
in the evolution kernel previously obtained in Ref.~\cite{Iancu:2015vea}. We are thus led
to the following, collinearly-improved, version of the BK equation,
 \begin{align}
 \label{colbk}
 	\frac{\dif \tilde{T}_{\bx\by}}{\dif Y} = 
 	\frac{\abar}{2\pi} \int \dif^2 \bz\,
 	\frac{(\bx \minus \by)^2}{(\bx \minus \bz)^2 
 	(\bz \minus \by)^2}\,
 	\left[\frac{(\bx \minus \by)^2}{\min\{(\bx \minus \bz)^2,(\by \minus \bz)^2\}}\right]^{\pm \abar A_1}&
 	\mathcal{K}_{\sdla}\big(\sqrt{L_{\bx\bz r}L_{\by \bz r}} \big)
 	\nn
 	&\hspace*{-1cm}\times \big( \tilde{T}_{\bx\bz} + \tilde{T}_{\bz\by} - \tilde{T}_{\bx\by} -
 	      \tilde{T}_{\bx\bz}\tilde{T}_{\bz\by}\big),
 \end{align}
 where the overall kernel is written as a product of three factors: the familiar dipole kernel which
 appears already at leading order, the `DLA kernel', resuming the double collinear logs to all orders \cite{Iancu:2015vea}
  \begin{equation}\label{kdla}
 	\mathcal{K}_{\sdla}(\rho) = \frac{\rmJ_1
 	\big(2\sqrt{\abar \rho^2}\big)}{\sqrt{\abar \rho^2}} = 
 	1- \frac{\abar \rho^2}{2} + \frac{(\abar\rho^2)^2}{12} + \cdots,
 \end{equation}
evaluated at $\rho=\sqrt{L_{\bx\bz r}L_{\by \bz r}}$, with $L_{\bx\bz r} \equiv \ln[(\bx-\bz)^2/r^2]$, and a new
factor, which features the exponent $\pm \abar A_1$ (the positive sign in the exponent is taken when 
$|\bx\minus\by|<\min\{|\bx \minus \bz|,|\by \minus \bz|\}$ and the negative sign otherwise), which 
expresses the contribution of the single collinear logarithms.

From the above discussion, is should also be clear that the present resummation of STLs
is only partial: it refers to the particular class of such corrections which
are generated by the first non-singular piece in the expansion in \eqn{pomega}.
The higher terms in this $\omega$--expansion will produce single collinear logarithms too, but only
starting at higher orders in perturbation theory (NNLO or higher). At the level of the BFKL
equation, more complete resummations of the single logarithms have been devised in
\cite{Salam:1998tj,Ciafaloni:1999yw,Ciafaloni:2003rd}, but so far it is not clear how to extend
these resummation schemes to a non-linear evolution equation like BK.

Returning to \eqn{colbk}, the tilde symbol in $\tilde{T}_{\bx\by}$ is intended
to remind that this is truly a suitable analytic continuation of the
dipole amplitude which
coincides with the physical quantity $T_{\bx\by}$ only for $\rho < Y$. For $\rho > Y$, the
physical amplitude can be obtained by either solving an equation non-local in $Y$, or by
matching onto the solution to the DGLAP equation \cite{Iancu:2015vea}. However,
explicit numerical studies at DLA level have shown that the solution $\tilde{T}_{\bx\by}$
to \eqn{colbk} remains very close to the actual physical amplitude, including for $\rho >Y$.
For this reason, we shall ignore this subtlety (and the related issue of the resummation
in the initial condition) for the purpose of the fits to be constructed
in the next section. We shall return to a more detailed study of these issues in
a forthcoming publication \cite{prep}.

\section{Prescriptions for the running of the coupling}
\label{sect:RC}

\begin{figure}[t]
\begin{minipage}[b]{0.33\textwidth}
\begin{center}
\includegraphics[scale=0.34]{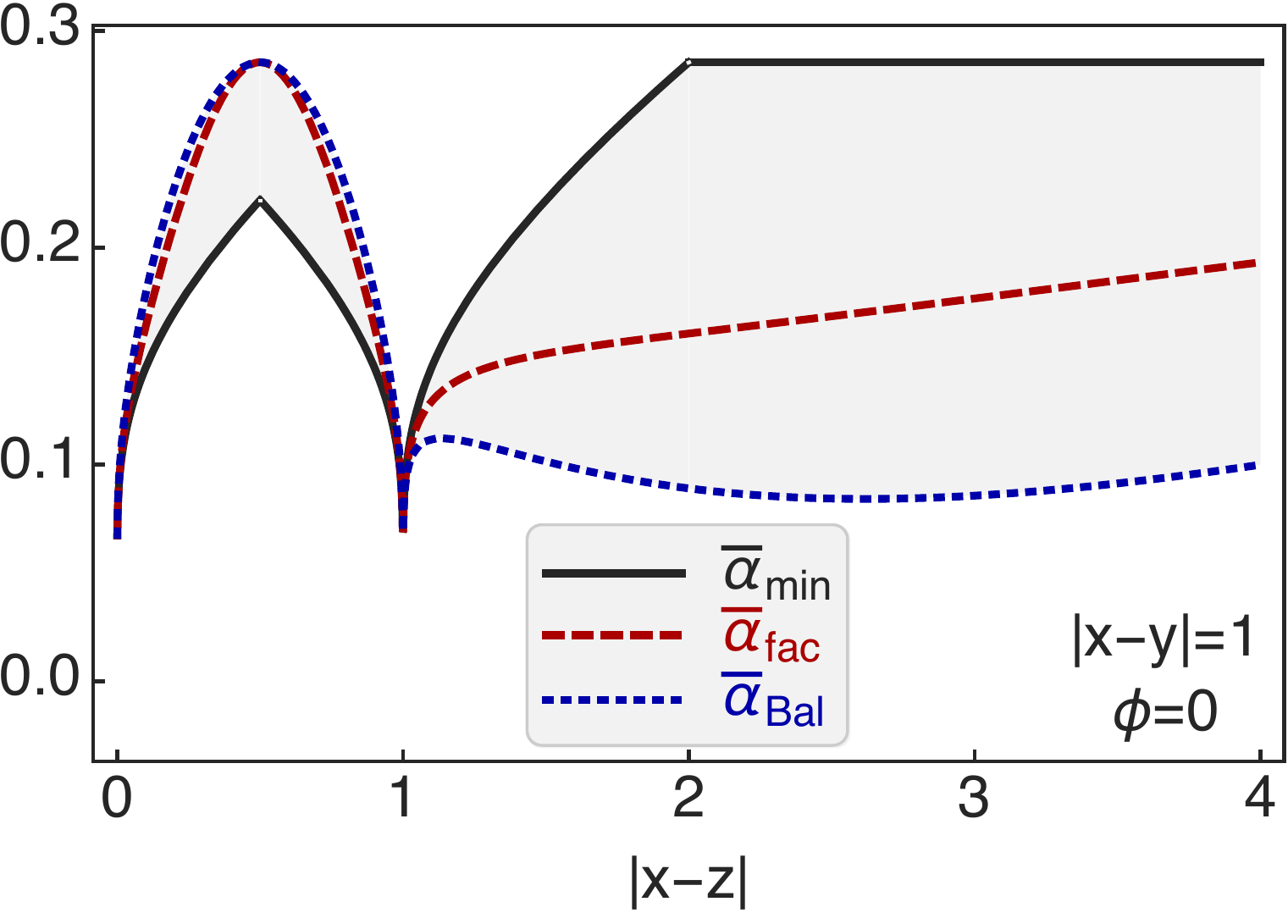}\\{\small (a)}
\end{center}
\end{minipage}
\begin{minipage}[b]{0.33\textwidth}
\begin{center}
\includegraphics[scale=0.34]{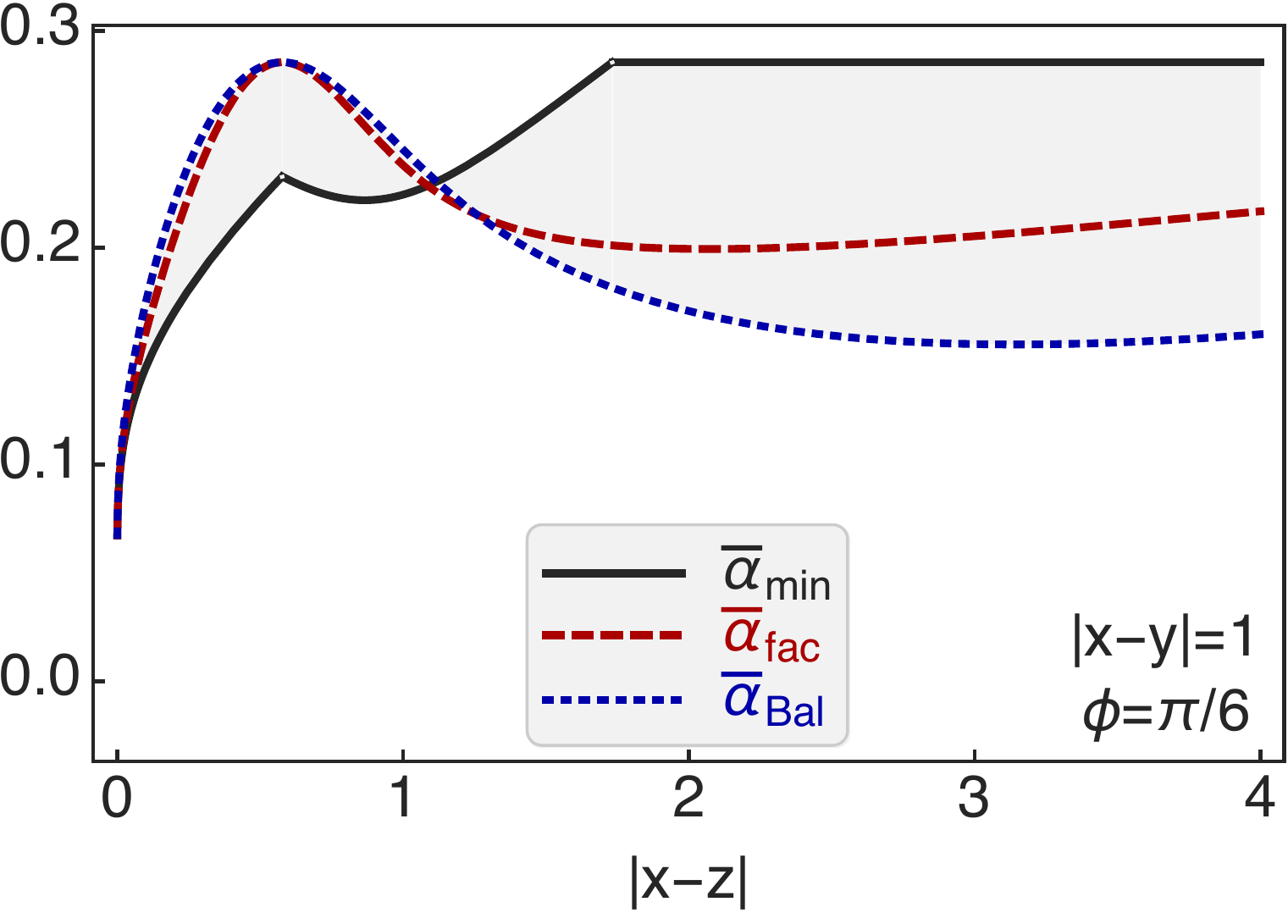}\\{\small (b)}
\end{center}
\end{minipage}
\begin{minipage}[b]{0.33\textwidth}
\begin{center}
\includegraphics[scale=0.34]{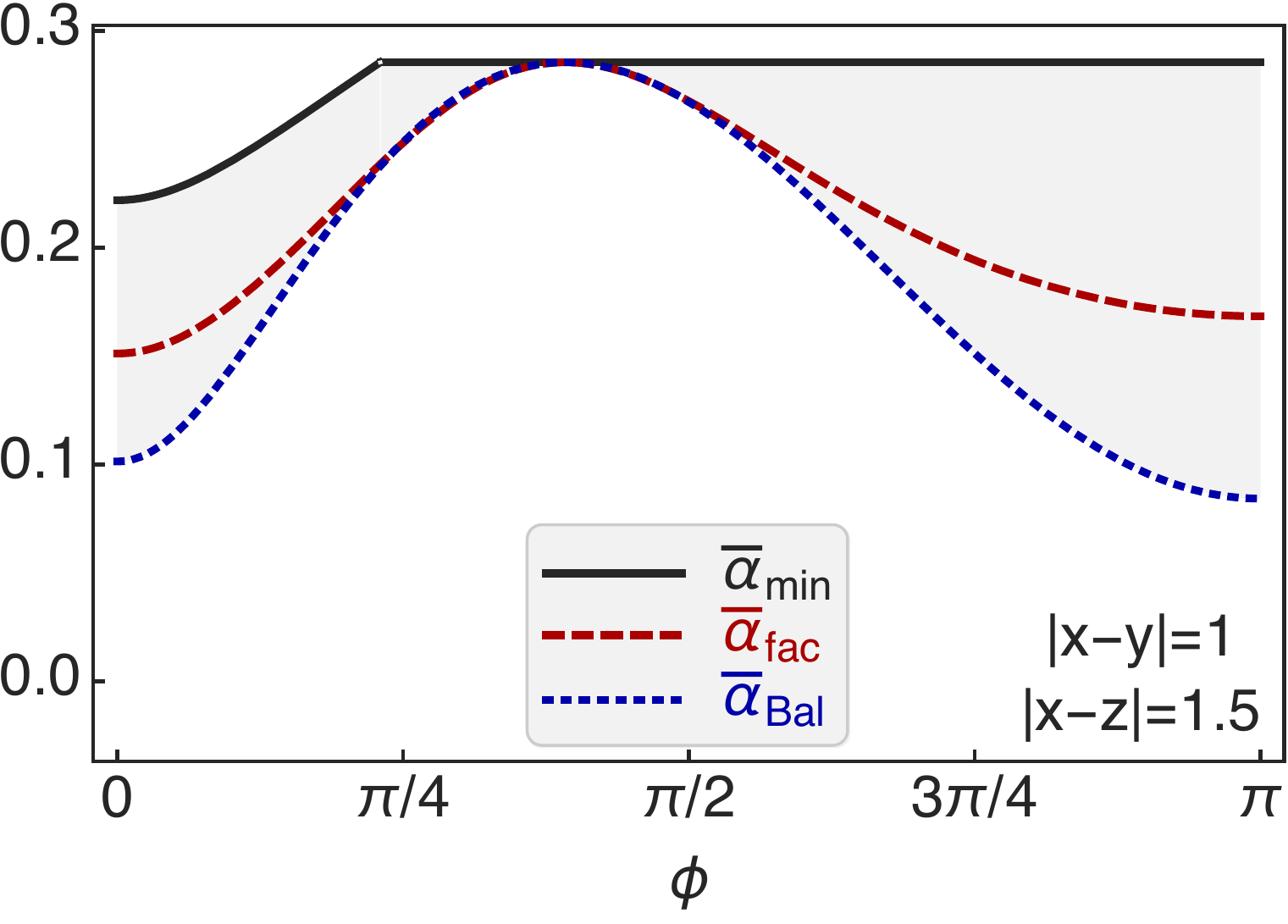}\\{\small (c)}
\end{center}
\end{minipage}
\caption{Running coupling for various schemes and configurations. (a)
  As a function of the daughter dipole size $|\bx-\bz|$, with
  $\phi=0$ the angle between the parent dipole $\bx-\by$ and the
  daughter one $\bx-\bz$. (b) The same with $\phi = \pi/6$. (c) As
  a function of the angle $\phi$ for fixed daughter dipole size
  $|\bx-\bz|=1.5$. Black (continuous): The minimal dipole scheme as
  defined in \eqn{amin}. Red (dashed): The ``fac'' scheme as given in
  \eqn{azero}. Blue (dotted): The Balitsky scheme
  \cite{Balitsky:2006wa}. In all cases the parent dipole size is
  $|\bx-\by|=1$, the coupling is smoothly frozen at the value 0.7 and
  $\Lambda_{\rm \scriptscriptstyle QCD} = 0.2$.}
\label{fig:run}
\end{figure} 

The last source of potentially large NLO corrections to the BK equation are the
running coupling corrections, i.e. the logarithmic terms proportional to $\bar{b}$ 
in the SI term in \eqn{nlobk}. Such terms can grow large
when the scales in their arguments are very disparate. More precisely,
the first logarithm can be problematic when $r$ is much smaller or
much larger than $1/\mu$, while the second when the soft gluon at
$\bz$ is collinear to the quark or the antiquark composing the parent
dipole. We need to choose $\mu$ in such a way to cancel these
potentially large logarithms, which could otherwise spoil the
convergence of the perturbative expansion\footnote{It is rather
  important to point out here that $\mu$ should cancel only these
  logarithms and not those discussed earlier which are of different
  physical origin. Of course one can proceed to such a choice and
  cancel all the NLO logarithms, but the result will be extremely
  unstable w.r.t.~small variations of $\mu$.}. It is clear that there
is not a unique choice, but in QCD one usually expects the hardest
scale to determine the running of the coupling. Indeed, a quick
inspection shows that the {\it smallest dipole prescription}
 \begin{equation}
 \label{amin}
 	\bar{\alpha}_{\rm min} = \abar(r_{\rm min})
 	\quad \mbox{with} \quad
 	r_{\rm min} = \min\{|\bx \minus\by|,|\bx \minus\bz|,|\by \minus\bz|\}  
 \end{equation}    
cancels the large logarithms in all kinematic regions.

Another possibility is to choose $\mu$ so that all the terms with
coefficient $\abar^2 \bar{b}$ vanish. Given that in the current work
we neglect all finite (i.e.~not enhanced by a large logarithm)
$\abar^2$ terms, this looks like what is called the {\it ``fastest
  apparent convergence'' (fac) scheme}
\cite{Grunberg:1980ja,Grunberg:1982fw,Ellis:1991qj}. It is convenient in
the sense that one is left with just the leading term in $\abar$. We
find that
 \begin{equation}
 \label{azero}
 	\bar{\alpha}_{\rm fac}
 = \left[\frac{1}{\abar(|\bx \minus \by|)} 
 + \frac{(\bx \minus \bz)^2 - 
 (\by \minus \bz)^2}{(\bx \minus \by)^2} \,
 \frac{\abar(|\bx \minus \bz|) - \abar(|\by \minus \bz|)}{\abar(|\bx \minus \bz|)\abar(|\by \minus \bz|)}\right]^{-1}, 
 \end{equation}      
and it is an easy exercise to show that it reduces to the minimal dipole choice $\abar(r_{\rm min})$ in all limits where one of the three dipoles is much smaller than the other two.

In this work, we shall use both above schemes. Let us add that the most popular prescription, widely used so far in phenomenological applications, is the one due to Balitsky \cite{Balitsky:2006wa}, and reads
\begin{equation}
 \label{abal}
 \bar{\alpha}_{\rm Bal}
 = \abar(|\bx \minus \by|)
 \left[1 + \frac{\abar(|\bx \minus \bz|) - 
 \abar(|\by \minus \bz|)}{\abar(|\bx \minus \bz|)\abar(|\by 
 \minus \bz|)}\,
 \frac{\abar(|\bx \minus \bz|) (\by \minus \bz)^2 - 
 \abar(|\by \minus \bz|) (\bx \minus \bz)^2}{(\bx \minus 
 \by)^2}
 \right], 
 \end{equation}
 but it will not be adopted here for a number of reasons.
 First, it is based on an extrapolation to all orders of a coordinate space kernel
 which includes the $\abar^2 \bar{b}$ terms above as well as $\abar^3
 \bar{b}^2$ corrections.
 At this order, we would also expect corrections proportional to the
 two-loop beta function.
 %
 %
 Second, even though it also reduces to $\abar(r_{\rm min})$ in the
 extreme kinematical limits, it does that very slowly for large
 daughter dipoles (in certain configurations) and this leads to an
 unphysically small coupling in a large region of phase space, as can
 be seen in the respective plots in Fig.~\ref{fig:run}. Finally, and
 perhaps as a result of the above drawbacks, when used in fitting the
 DIS data, it gives a much worse fit than Eqs.~\eqref{amin} and
 \eqref{azero} and with fit parameters which take somewhat unnatural
 values.

\section{Fits to the HERA data}
\label{sec:fits}

We now turn to the description of the HERA reduced cross-section
measurements using the resummed BK equation. To this aim several
ingredients first have to be specified.

\paragraph{Initial condition} We must fix the initial condition for
the dipole amplitude at some $Y_0$, which afterwards will be evolved
towards higher rapidities using \eqn{colbk}. We consider two choices:
the simple parametrisation of the Golec-Biernat and W\"{u}sthoff (GBW)
\cite{GolecBiernat:1998js} type
\begin{equation}
  \label{eq:gbwinit}
  T(Y_0,r)= 
  \left\{1 -\exp\left[-\left(\frac{r^2Q_0^2}{4}
  \right)^p\right]\right\}^{1/p}
\end{equation}
and the running-coupling version of the McLerran-Venugopalan (rcMV)
model \cite{McLerran:1993ka}
 \begin{equation}
 \label{eq:runinit}
  T(Y_0,r)=
  \left\{1-\exp\left[-\left(\frac{r^2Q_0^2}{4}\,
              \bar\alpha_s(r)\left[1+
                \ln\left(\frac{\bar\alpha_{\rm sat}}
                               {\bar\alpha_s(r)}
              \right)\right]\right)^p\right]\right\}^{1/p}.
\end{equation}
It is worth noticing that, as dictated by collinear physics, there is
no anomalous dimension in the above initial conditions. The extra
parameter $p$ determines the shape of the amplitude close to
saturation and its approach towards unitarity. 

\paragraph{Running coupling} We consider the two prescriptions
given by Eqs.~\eqref{amin} and
\eqref{azero}. For the explicit expression of the strong coupling in
coordinate space in terms of $r$
we introduce a fudge factor as in \cite{Albacete:2010sy}, namely
 \begin{equation}
 \alpha_s(r) =
 \frac{1}{b_{\Nf}\ln\big[4C_\alpha^2/(r^2\Lambda_{\Nf}^2)\big]},
 \end{equation}
 with $b_{\Nf}=(11 \Nc- 2 \Nf)/12\pi$. This fudge factor is also
 included in the rcMV type initial condition
 in~(\ref{eq:runinit}). The $\Nf$-dependent Landau pole is obtained by
 imposing $\alpha_s(M_Z^2)=0.1185$ at the scale of the $Z$ mass
 \cite{Agashe:2014kda} and continuity of $\alpha_s$ at the flavour
 thresholds, using $m_c=1.3$~GeV and $m_b=4.5$~GeV. To regularise the
 infrared behaviour, we have decided to freeze $\alpha_s$ at a value
 $\alpha_{\rm sat}=1$ and we have checked explicitly that reducing
 this down to, for example, 0.7 does not affect the fit in any
 significant manner.

 Note that we do not include any form of resummation or matching for
 $\ln 1/r^2 > Y$, as introduced in \cite{Iancu:2015vea}, in these
 initial conditions. One of the reasons for not doing so is that the
 extra factor in the initial condition can always be reabsorbed in a
 re-parametrisation. Furthermore, a proper matching at small dipole
 sizes, suited for phenomenological studies, would require a careful
 treatment of the small-dipole region. In that respect, the resummed
 BK evolution is expected to perform a better job than a fixed
 matching with a fixed asymptotic behaviour. We leave a better treatment,
 e.g. a genuine matching to DGLAP evolution, 
 for future work.

 \paragraph{Rapidity evolution} Of course this is determined by the
 resummed BK equation given in \eqref{colbk}. Here, we again consider
 two separate cases, one in which the evolution resums only the
 leading double logarithms and one in which it also includes the
 single ones.

\begin{table}
  \begin{center}
    \begin{tabular}{|l|l|l|c|c|c|c|c|c|c|}
      \hline
      init & RC & sing. & \multicolumn{3}{c|}{$\chi^2$ per data point} 
      & \multicolumn{4}{c|}{parameters} \\
      \cline{4-10}
      cdt. & schm & logs & $\sigma_{\rm red}$ & $\sigma_{\rm red}^{c\bar c}$ & $F_L$
                  & $R_p$[fm] & $Q_0$[GeV] & $C_\alpha$ &  $p$ \\
      \hline
      GBW  & small & yes & 1.135 & 0.552 & 0.596 & 0.699 & 0.428 & 2.358 & 2.802 \\
      GBW  & fac   & yes & 1.262 & 0.626 & 0.602 & 0.671 & 0.460 & 0.479 & 1.148 \\
      rcMV & small & yes & 1.126 & 0.565 & 0.592 & 0.707 & 0.633 & 2.586 & 0.807 \\
      rcMV & fac   & yes & 1.228 & 0.647 & 0.594 & 0.677 & 0.621 & 0.504 & 0.541 \\
      GBW  & small & no  & 1.121 & 0.597 & 0.597 & 0.716 & 0.414 & 6.428 & 4.000 \\
      GBW  & fac   & no  & 1.164 & 0.609 & 0.594 & 0.697 & 0.429 & 1.195 & 4.000 \\
      rcMV & small & no  & 1.093 & 0.539 & 0.594 & 0.718 & 0.647 & 7.012 & 1.061 \\
      rcMV & fac   & no  & 1.132 & 0.550 & 0.591 & 0.699 & 0.604 & 1.295 & 0.820 \\
      \hline
    \end{tabular}
  \end{center}
  \caption{$\chi^2$ and values of the fitted parameters entering the description of the HERA data. 
    The fit includes the 252 $\sigma_{\rm red}$ data points. The
    quoted $\chi^2$ for $\sigma_{\rm red}^{c\bar c}$ and $F_L$ are
    obtained a posteriori.}
    \label{tab:parameters}
\end{table}

\paragraph{From the dipole amplitude to observables}

Once we have the dipole amplitude for all rapidities and dipole
sizes, we use the standard dipole formalism to obtain the physical
observables:
 \begin{equation}
 \label{eq:dipole-to-sigma}
 \sigma_{\rm L,T}^{\gamma^*p}(Q^2,x) = 
 2\pi R_p^2\,\sum_f\int {\rm d}^2r\int_0^1
 {\rm d}z\,\big|\Psi_{\rm L,T}^{(f)}(r,z;Q^2)\big|^2\,
 T(\ln 1/\tilde x_f,r),
\end{equation}
 where the transverse and longitudinal virtual photon wavefunctions read
 \begin{align}
 \label{eq:wave-functions}
 \big|\Psi_{\rm L}^{(f)}(r,z;Q^2)\big|^2 & = 
 e_q^2\frac{\alpha_{\rm em}N_c}{2\pi^2}\,
 4Q^2z^2(1-z)^2K_0^2(r\bar Q_f),\\
 \big|\Psi_{\rm T}^{(f)}(r,z;Q^2)\big|^2 & = 
 e_q^2\frac{\alpha_{\rm em}N_c}{2\pi^2}\,
 \left\{\left[z^2+(1-z)^2\right]\bar Q_f^2 K_1^2(r\bar Q_f)
 + m_f^2 K_0^2(r\bar Q_f)\right\}.
 \end{align}
 In the above we have introduced the customary notation $\bar
 Q_f^2=z(1-z)Q^2+m_f^2$, $\tilde x_f=x(1+4m_f^2/Q^2)$, and we have
 assumed a uniform distribution over a disk of radius $R_p$ in impact
 parameter space. 
 The sum in~(\ref{eq:dipole-to-sigma}) runs over all quark flavours
 and we will include the contributions from light quarks with
 $m_{u,d,s}=100$~MeV as well as from the charm quark with $m_c=1.3$~GeV.
 From the longitudinal and transverse cross-sections,
 we can deduce the reduced cross-section and the longitudinal
 structure function as
 \begin{align}
 \label{eq:sigmared-FL}
 \sigma_{\rm red} & = \frac{Q^2}{4\pi^2\alpha_{\rm em}}
 \left[\sigma_{\rm T}^{\gamma^*p}+
 \frac{2(1-y)}{1+(1-y)^2}\sigma_{\rm L}^{\gamma^*p}\right],
 \\
 F_{\rm L} & = \frac{Q^2}{4\pi^2\alpha_{\rm em}}
 \sigma_{\rm L}^{\gamma^*p}.
 \end{align}

 When the quark masses, the value of the strong coupling at the $Z$
 mass and its frozen value in the infrared have been fixed, we are
 left with 4 free parameters according to our choice of initial
 condition: $R_p$ the ``proton radius'', $Q_0$ the scale separating
 the dilute and dense regimes, $C_\alpha$ the fudge factor in the
 running coupling in coordinate space, and $p$ which controls the
 approach to saturation in the initial condition.

 We have fitted these parameters to the combined HERA measurements of
 the reduced photon-proton cross-section \cite{Aaron:2009aa}. Since
 the BK equation is applicable only at small-$x$, we have limited
 ourselves to the region $x \le 0.01$. We note that since
 \eqn{eq:dipole-to-sigma} probes dipoles at the rapidity $\ln 1/\tilde
 x_f$, the exact cut we impose is $\tilde{x}_c \le 0.01$ since the
 most constraining cut comes from the charm, the most massive quark we
 include in our model. Accordingly, our initial condition for the BK
 evolution corresponds to $\tilde{x}=0.01$. Furthermore, since we do
 not expect the BK equation to capture the full collinear physics, we
 impose the upper bound $Q^2<Q^2_{\rm max}$. By default we will use
 $Q^2_{\rm max}=50$~GeV$^2$ but we will also give results for
 extensions to larger $Q^2$. In the default case we have a total of
 252 points included in the fit.
 We have added the statistical and systematic uncertainties in
 quadrature.\footnote{A more involved treatment of the correlated
   systematic uncertainties leads to similar results with slightly
   worse $\chi^2$ per points (about 0.04).}

 The results of our fits for the $2^3=8$ cases, depending on the
 initial condition, the running coupling prescription and the inclusion or
 not of single logarithms in the kernel, are presented in
 Table.~\ref{tab:parameters}.
 The table includes the parameter values obtained from fitting the
 $\sigma_{\rm red}$ data and, besides the fit $\chi^2$, it also
 indicates the $\chi^2$ obtained a posteriori for the latest
 $\sigma_{\rm red}^{c\bar c}$ \cite{Aaron:2009af} and $F_L$
 \cite{Collaboration:2010ry} measurements.
 These results deserve a few important comments.

\begin{table}
  \begin{center}
    \begin{tabular}{|l|l|l|c|c|c|c|}
      \hline
      init & RC & sing. & \multicolumn{4}{c|}{$\chi^2/\text{npts}$ for $Q^2_{\rm max}$} \\
      \cline{4-7}
      cdt. & schm & logs &   50  &  100  &  200  &  400  \\
      \hline
      GBW  & small & yes & 1.135 & 1.172 & 1.355 & 1.537 \\
      GBW  & fac   & yes & 1.262 & 1.360 & 1.654 & 1.899 \\
      rcMV & small & yes & 1.126 & 1.170 & 1.182 & 1.197 \\
      rcMV & fac   & yes & 1.228 & 1.304 & 1.377 & 1.421 \\
      GBW  & small & no  & 1.121 & 1.131 & 1.317 & 1.487 \\
      GBW  & fac   & no  & 1.164 & 1.203 & 1.421 & 1.622 \\
      rcMV & small & no  & 1.093 & 1.116 & 1.106 & 1.109 \\
      rcMV & fac   & no  & 1.131 & 1.181 & 1.171 & 1.171 \\
      \hline
    \end{tabular}
  \end{center}
  \caption{Evolution of the fit quality when including
    data at larger $Q^2$ (in GeV$^2$). 
}
\label{tab:largeQ2}
\end{table}

\begin{figure}
  \centerline{\includegraphics[angle=0,width=0.75\textwidth]{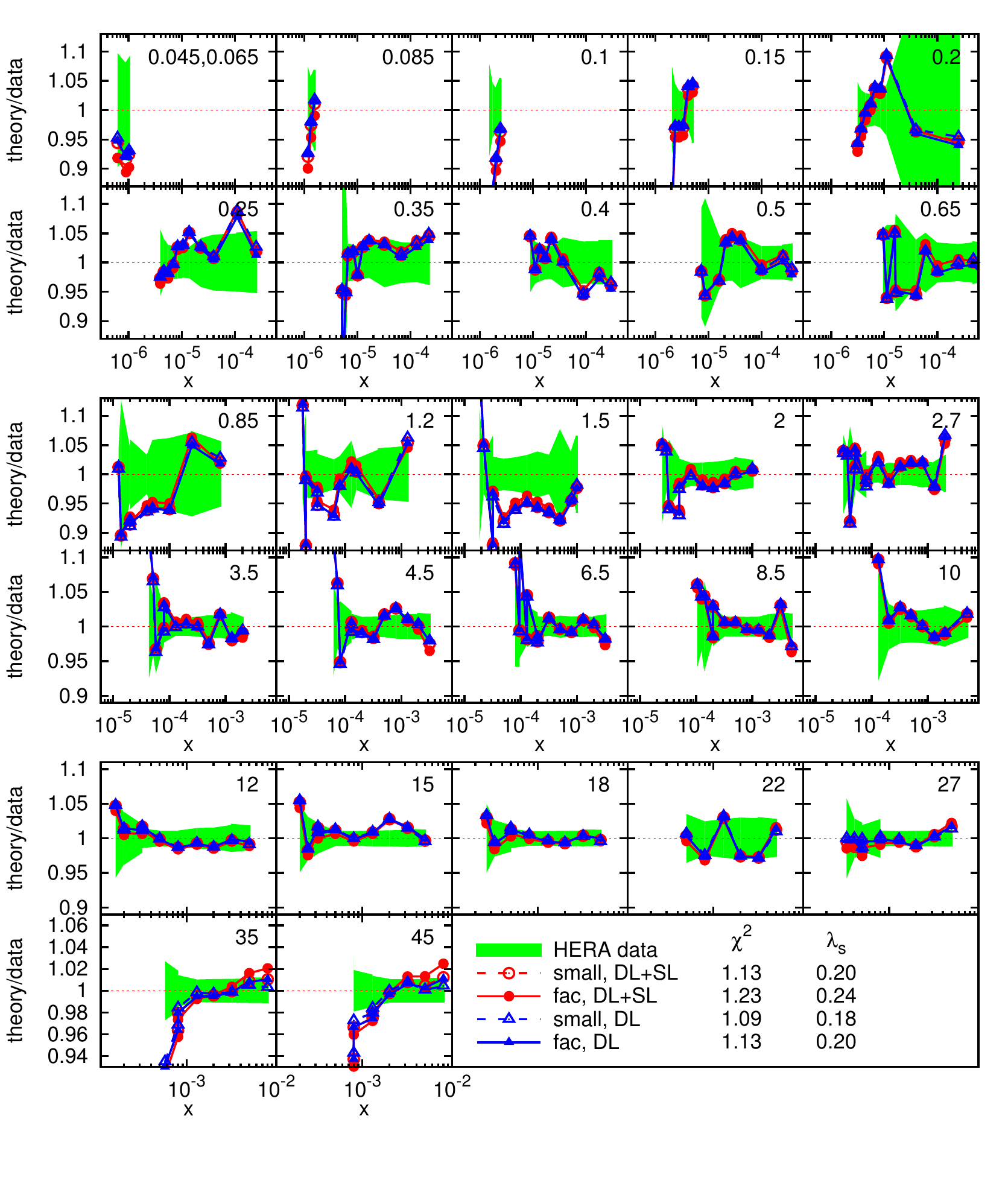}}
\caption{Description of the HERA data obtained by the fits using the
  rcMV initial condition. Each box corresponds to a given value of
  $Q^2$ as indicated (in GeV$^2$) in the top-right corner. For each fit
  we plot the ratio of the prediction to the central experimental
  value. The (green) band represents the experimental uncertainty.}
  \label{fig:fit-with-data}
\end{figure}

\begin{enumerate}
\item In general, the overall quality of the fit is very good,
  reaching $\chi^2$ per point around 1.1-1.2.
\item Apart from a few small exceptions (see below), all the parameters
  take acceptable values of order one. Note that we have manually
  bounded $p$ between 0.25 and 4. Whenever it reached the
  upper limit, larger values only led to minor improvements in the
  quality of the fit.
\item The two initial conditions give similar results, with a slight
  advantage for the rcMV option, which is likely due to the extra
  parameter.
  Note that for
  a standard MV-type of initial condition 
  $ T(Y_0,r)=\{1-\exp[-(r^2Q_0^2/4\,[c+\ln(1+1/r\Lambda)])^p]\}^{1/p}$,
  we have not been able to obtain a $\chi^2$ per point below 1.3 and
  the parameters, typically $c$ or $p$, tend to take unnatural
  values.
\item As far as the running-coupling prescription is concerned, the
  smallest dipole prescription given in \eqn{amin} tends to give
  somewhat better fits than the ``fac'' prescription given in
  \eqn{azero}. 
  This can be seen as an estimate of subleading corrections (including
  the pure $\abar^2$ NLO terms) that we neglect in the present fit.
  Note also that we have not been able to reach a fit of equivalent
  quality and robustness with the Balitsky prescription.
\item The resummation of the single logarithms tends to yield slightly
  larger values for $\chi^2$, but the difference is too small to be
  significative (at least, without performing a full NLO analysis).
  Perhaps more significantly, this resummation leads to more physical
  values for some of the parameters, especially $C_\alpha$ for the
  smallest dipole prescription and $p$ for the GBW initial condition.
  %
  %
  These findings are consistent with the expectation that, once
  properly resummed, single logarithms should have only a modest
  impact.
  Recall however that their resummation is a crucial step
  towards a full NLO fit --- failing to do so could lead to
  instabilities similar to those observed when double logarithms are
  not resummed.
\item The fit remains stable when varying the parameters we have
  imposed by hand. For example, using $\alpha_{\rm sat}=0.7$ instead
  of $1$ has no significant effect on the fit. Varying the light quark
  masses within the rather wide range $0\le m_{u,d,s}\le 140$~MeV
  only slightly changes the quality of the fit. For instance, taking one of our best fits
  (rcMV initial condition, the smallest dipole prescription for the running
  of the coupling, and resummation of the single logarithms),
   we have found $\chi^2=\{1.180,\,1.153,\,1.126,\,1.159\}$
  when choosing $m_{u,d,s}=\{0,\,50,\,100,\,140\}$~MeV, respectively.
  This lack of sensitivity to the light quark masses is likely a consequence 
  of saturation, which reduces the dependence of the DIS  
  cross-section to very large dipole fluctuations. (The corresponding amplitudes
  reach the unitarity, or  `black disk', limit $T=1$, so they are independent of
  the size $r$ of the dipoles fluctuations, as regulated at low $Q^2$
  by the quark masses; see also the discussion of Fig.~\ref{fig:Qs} below.)
  Also, we have obtained an equally good fit with the slightly larger value
  $m_c=1.4$~GeV for the mass of the charm quark, although the quality
  started deteriorating for significantly  larger values  $m_c\ge 1.6$~GeV.
  Very similar findings have been reported for the saturation fits in  \cite{Rezaeian:2012ji}.

\item Trying to extend the fit to larger $Q^2$ shows an interesting
  behaviour as seen from Table~\ref{tab:largeQ2}. While the $\chi^2$
  obtained using the GBW initial condition increase when including
  higher-$Q^2$ data, the fits using the rcMV initial condition remain
  stable. We suspect that this is due to the fact that this choice of
  initial condition stays closer to the expected physics at high
  $Q^2$.
\end{enumerate}

In Fig.~\ref{fig:fit-with-data} one can see the quality of our fit and
the extracted values of the evolution parameter $\lambda_s = \dif \ln
Q_s^2/\dif Y$. In Fig.~\ref{fig:Qs} we show the value of the
saturation momentum in the ($x,Q^2$)-plane on top of the data points
as well as a few selected initial conditions for the fit. Note that amplitudes 
which a priori have different functional forms, cf. Eqs.~\eqref{eq:gbwinit}
and \eqref{eq:runinit}, look nevertheless quite similar in shape
(at least in double-logarithmic scale) when plotted for the particular
values of the parameters that are selected by the fits.

\begin{figure}
  \centerline{\includegraphics[angle=0,height=7.cm]{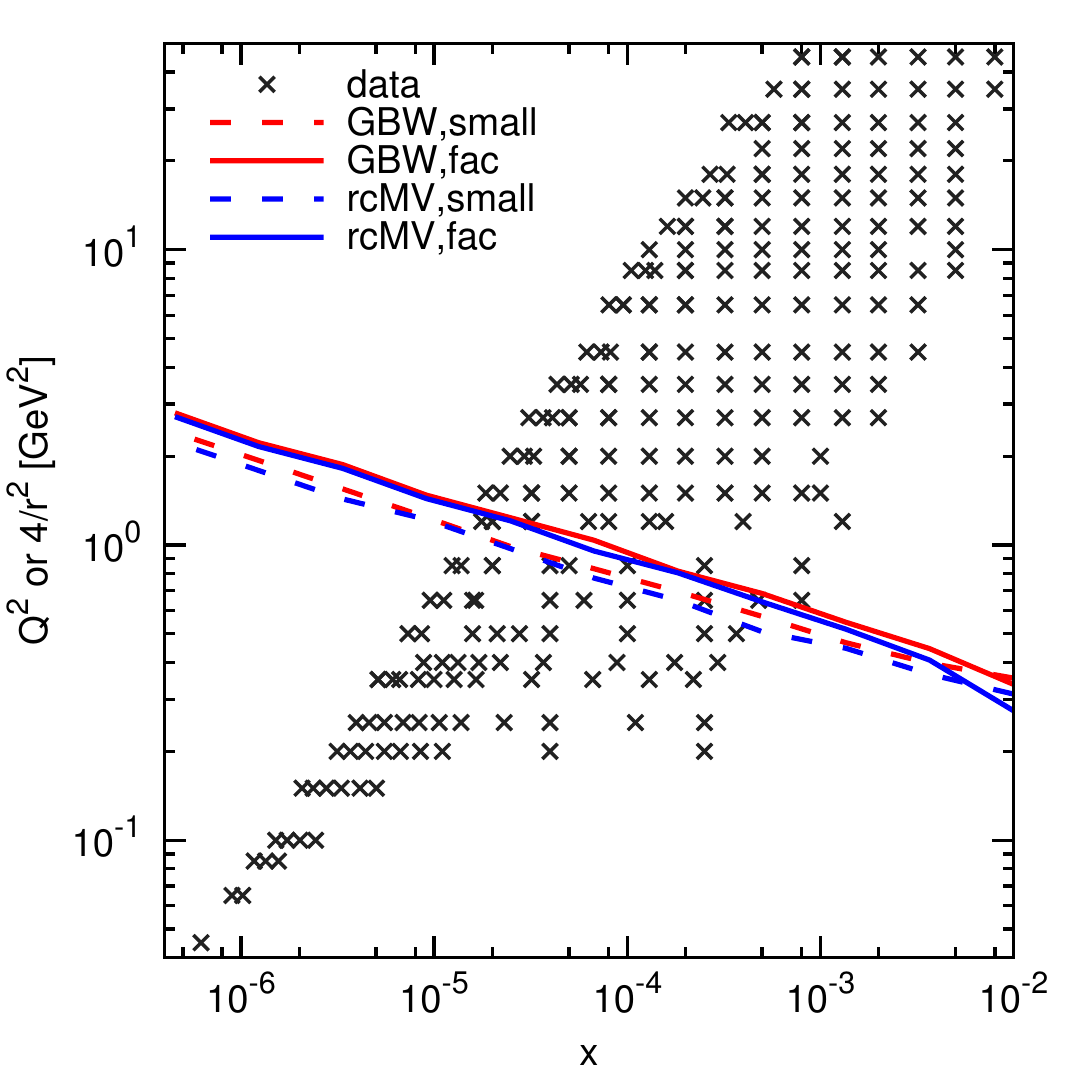}
  \hfill\includegraphics[angle=0,height=7.1cm]{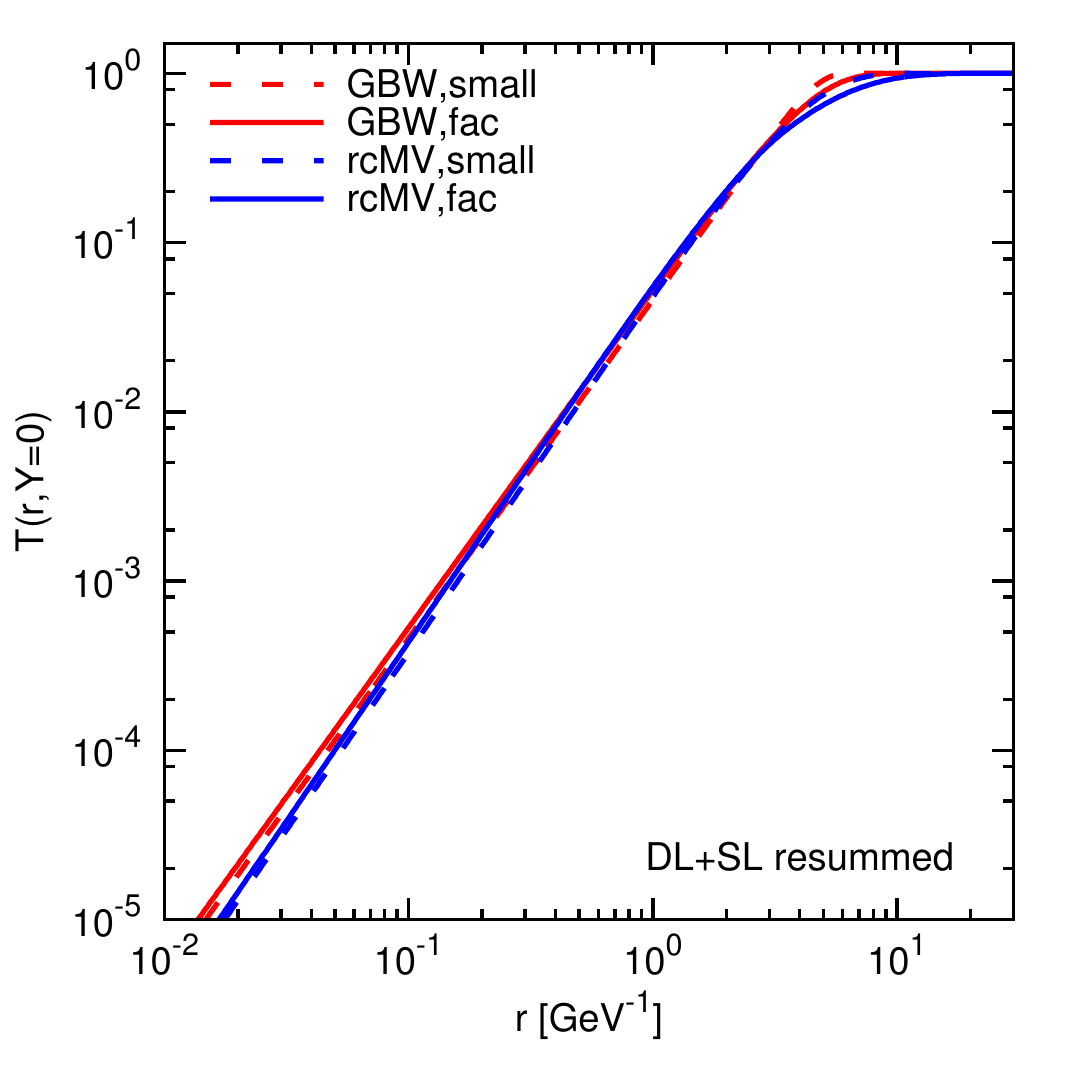}
  }
  \caption{Left: value of the saturation momentum, defined for each
    rapidity as $2/r_s(Y)$ with $T(r_s(Y), Y)=1/2$. For comparison, we
    have overlaid the experimental data points from the HERA
    dataset. Right: plot of the corresponding initial conditions for
    the rapidity evolution.}
  \label{fig:Qs}
\end{figure}

To conclude, this work can be seen as the first description of
small-$x$ DIS data which includes mandatory perturbative QCD
ingredients in that region: leading-order small-$x$ evolution, the
resummation of large transverse logarithms, and saturation
corrections\footnote{It would be an interesting exercise to see what
  happens if one switches off the non-linear corrections in
  Eq.~(\ref{colbk}). Given our asymmetric choice of frame, justified
  by saturation physics, this may however require extra work. See also
  \cite{Hentschinski:2013id}.}.
The dipole amplitude obtained from our fits to inclusive DIS 
can in principle be used to compute several other observables, 
like particle multiplicity in hadronic collisions,
the diffractive structure functions, the elastic production of vector
mesons, or the forward particle production in heavy-ion collisions.
This is certainly not the end of the story: beyond what we have
included here, it would be interesting to add the pure $\abar^2$ NLO
corrections to the BK evolution kernel, thus obtaining a genuine
resummed NLO-BK fit, and to perform a proper matching between this
small-$x$ evolution and a DGLAP-like evolution at large $Q^2$ and
large $x$.
These steps go beyond the scope of the present paper and are left for
future studies.
      
\section*{Acknowledgements}
This work is supported by the European Research Council under the Advanced Investigator Grant ERC-AD-267258 and by the Agence Nationale de la Recherche project \# 11-BS04-015-01.  The work of A.H.M.
is supported in part by the U.S. Department of Energy Grant \# DE-FG02-92ER40699. 
G.S.~wishes to thank Javier Albacete and Guilherme Milhano for helpful discussions on the AAMQS results.

\bibliographystyle{utcaps}
\bibliography{refs}

\end{document}